%% file: ms.tex
\documentclass[usenames,dvipsnames]{IEEEtran}
\usepackage{graphicx}
\usepackage{ifpdf}
\ifpdf
   \DeclareGraphicsExtensions{.pdf,.png,.jpg,.mps}
   \usepackage{hyperref} 
\else
\fi

\usepackage{mystyle}
\makeatletter
\def\ps@headings{%
\def\@oddhead{\mbox{}\scriptsize\rightmark \hfil \thepage}%
\def\@evenhead{\scriptsize\thepage \hfil \leftmark\mbox{}}%
\def\@oddfoot{}%
\def\@evenfoot{}}
\makeatother
\begin{document}

\title{Design of a Hybrid Modular Switch}
\author{\IEEEauthorblockN{Ashkan Aghdai,\, Yang Xu,\, H. Jonathan Chao}
\IEEEauthorblockA{\\New York University\\Tandon School of Engineering \\ \{ashkan.aghdai,\ yang,\ chao\}@nyu.edu}}

%\IEEEpeerreviewmaketitle
\maketitle
\thispagestyle{empty}

\begin{abstract}
    Network Function Virtualization (NFV) shed new light for the design, deployment, and management of cloud networks.
    Many network functions such as firewalls, load balancers, and intrusion detection systems can be virtualized by servers.
    However, network operators often have to sacrifice programmability in order to achieve high throughput, especially at networks' edge where complex network functions are required.

    Here, we design, implement, and evaluate Hybrid Modular Switch (HyMoS).
    The hybrid hardware/software switch is designed to meet requirements for modern-day NFV applications in providing high-throughput, with a high degree of programmability.
    HyMoS utilizes P4-compatible Network Interface Cards (NICs), PCI Express interface and CPU to act as line cards, switch fabric, and fabric controller respectively.
    In our implementation of HyMos, PCI Express interface is turned into a non-blocking switch fabric with a throughput of hundreds of Gigabits per second.
    
    Compared to existing NFV infrastructure, HyMoS offers modularity in hardware and software as well as a higher degree of programmability by supporting a superset of P4 language.
\end{abstract}


    \input{intro.tex}
    \input{architecture.tex}

    \input{challenges.tex}

    \input{evaluation.tex}

    \input{related.tex}
    \input{future.tex}

    \bibliographystyle{ieeetr}
    \tiny{\bibliography{ms.bib}}

\end{document}

%% file: intro.tex
\section{Introduction}
Network functions play a significant role in shaping, policying, and monitoring the Internet traffic.
%For instance, stateful firewalls and intrusion detection systems provide scalable security and isolation services.
%Cloud operating systems such as OpenStack \cite{openstack} utilize middleboxes' raw packet processing power to implement networking services.
%However, fixed-function hardware devices usually come short when customized functionalities are required.
%
Network Function Virtualization (NFV) lets ISP and Cloud operators utilize programmable devices to tailor the data plane behavior according to their needs. 
Recent advances in Software Defined Networks (SDN) provides a foundation for building programmable networks.
OpenFlow~\cite{mckeown2008openflow} started a new trend in network design and operation by isolating the control plane from the data plane, and Network Operating System (NOX)~\cite{gude2008nox} allows operators to build applications on top of programmable hardware or software OpenFlow-enabled switches.
In addition to providing a solid platform for virtualization of network functions, a substantial number of innovative network applications are developed using SDN.
Smart rule caching and placement \cite{yu2010scalable,katta2014infinite,yan2014cab}, intelligent access control \cite{kim2013improving,porras2012security}, policy verification \cite{kazemian2013real,khurshid2012veriflow}, high-level network programming languages \cite{foster2011frenetic,reich2013modular}, and advanced network measurements tools \cite{yu2013software,moshref2014dream} are just a few examples of SDN applications that enhance performance, manageability, and flexibility of networks.

Considering the large body of work in developing applications and implementing new ideas in SDN over the years, it is surprising that design and implementation of high-performance programmable switches, devices that enables all SDN applications, has not received as much attention before the introduction of Protocol-independent Switch Architecture (PISA)~\cite{bosshart2013forwarding}.
PISA proposes Reconfigurable Match Tables (RMT), a key paradigm shift in designing programmable switches.
RMT introduces a generalized high-performance processing model and redefines packet forwarding as a domain problem~\cite{van2000domain}.
P4 language~\cite{bosshart2014p4}, as it stands for Programmable Protocol-independent Packet Processing, introduces a much needed Domain Specific Language (DSL)~\cite{van2000domain} for programmable packet forwarding.
As opposed to OpenFlow-like protocols that aim at providing a reliable means for distribution and management of forwarding rules, P4 exposes the inner workings of programmable switches;
it allows users to identify packet headers using a programmable parser, specify the matching fields as well as a set of available actions for each forwarding table, and lay out the flow of packets between match+action tables.
PISA and PISCES~\cite{shahbaz2016pisces} are examples of P4 targets i.e. devices that can be reconfigured with a P4 program.
PISCES is a compatibility layer on top of Open vSwitch (OVS)~\cite{pfaff2015design} that relies on software to process packets according to a P4 program.
%PISA and PISCES are the two ends of packet processing spectrum, pure hardware-based and pure software-based packet processing respectively.
%
%\section{Case for Programmabality Beyond P4}

P4 introduces a target-independent language for programmable devices, but it lacks some much needed features for NFV applications.
%For instance, the structure of P4 programs is tightly coupled to PISA architecture and it confines programmers to describe data plane functionality for fixed ingress/egress pipeline.
%The language does not support negative matching.
%While it standardizes the switch configuration, it does not define a northbound control protocol to populate custom tables.%, and its introduction of programmable parsers makes existing protocols such as OpenFlow obsolete.
For instance, P4 cannot program switch scheduler, manage queues, or process packets beyond pre-defined header fields.
Such limitations effectively bar developers from implementing QoS protocols, or more advanced packet processing techniques such as Deep Packet Inspection (DPI), common functions in networks' edge. 
%packet processing is limited on headers only, deep packet inspection (DPI) mechanisms.%, and complicated packet parsing techniques such as parsers controlled by the control flow.
%Therefore the language is geared towards programming high-throughput L2 and L3 core networks and it is less suitable for deploying advanced L4 and up virtual functions at networks’ edge that usually run at a slower pace compared to core networks.
As a result, more flexible architectures such as NetVM~\cite{hwang2015netvm} and Open Compute Project (OCP) Wedge~\cite{ocp2015wedge} rely on Intel Data Plane Development Kit (DPDK)~\cite{intel2014data} and X86-powered processing to implement a highly customizable data plane.
Internet Service Providers (ISP) and data center operators alike use such designs to deploy NFV applications at networks’ edge where a higher degree of programmability is required at $O(100G)$ or higher aggregated bandwidth~\cite{att2016ecomp,han2015network,att2013domain}.
However, DPDK's improved programmability comes at the cost of limited packet processing power on X86 CPUs.

To address the gap between NFV applications' requirements at networks’ edge and available solutions, we take a fresh approach to design a new family of programmable switches.
We aim at bringing the best of the hardware and software switches to design a highly programmable Hybrid Modular Switch (HyMoS) with the following objectives in mind:

\begin{itemize}
    \item Aggregated throughput suited for networks’ edge, i.e., dense 10/40G switches.
    \item Programmability beyond P4 language.
    \item QoS capabilities that enable network operators to define and meet Service Level Agreements (SLAs).
    \item Modularity in terms of hardware and software.
%        Switch operators should be able to not only mix line cards according to port count/rate requirements but also install and develop switch-level agents for SDN applications or NFV functions.
\end{itemize}

We have designed, prototyped, and evaluated the performance of HyMoS.
Our early results show that it is possible to build a cost-effective switch based on commodity servers that utilizes P4-compatible Network Interface Cards (NIC) and PCI Express (PCI-e)~\cite{pci2010rev3} backplane to process and switch packets, respectively.

The rest of the paper is organized as follows.
Section~\ref{architecture} introduces HyMoS' architecture.
%\autoref{challenges} presents the design challenges.
Section~\ref{evaluation} evaluates HyMoS and presents early results.
Section~\ref{related} reviews related works in this area.
Finally, Section~\ref{future} concludes the paper.

%% file: architecture.tex
\section{A Radical Switch Design Approach}\label{architecture}

\begin{figure}[t]
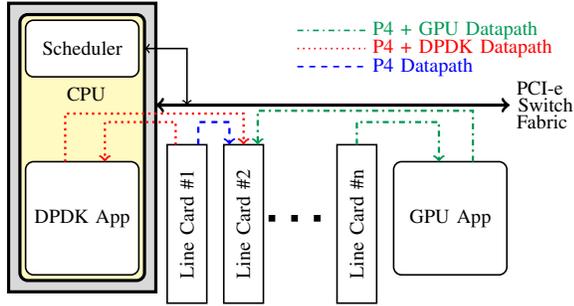

    \centering
    \includestandalone[width=0.425\textwidth]{MoPPetS-Arch}
    \caption{Proposed switch architecture}
    \label{HyMoS-Arch}
\end{figure}

Network devices perform two operations on every packet: processing and switching.
Packet processing involves table lookups on specific fields in the packet header.
It also decides which output ports - if any - the packet should be forwarded to.
Therefore, depending on the active networking protocols, packet processing can be very complex, and its software implementations' throughput is usually CPU-bound \cite{shahbaz2016pisces}.
Packet switching, however, only involves copying processed packets from ingress port to specified egress port/s.

To achieve high throughput, HyMoS relies on line cards to process packets and perform destination look ups using hardware.
Due to the simplicity of switching processed packets, we rely on X86 processors to orchestrate memory operations between ingress and egress ports to switch packets.
We show that underlying switch fabric has enough available throughput to perform the required operations.

\subsection{HyMoS' Architecture}
HyMoS utilizes Network Interface Cards (NIC) as line cards, PCI-e interface as the switch fabric, and CPU as the scheduler to build a modular switch.
NICs process packets at hardware and perform table lookups.
Once processed, packets are stored in a Virtual Output Queue (VOQ)~\cite{mckeown1999achieving} structure according to their destination line cards.
A pipelined CPU process polls VOQs, arbitrates between the requests, and schedules peer-to-peer PCI-e transactions for granted requests.
The small size of the corresponding bipartite matching~\cite{west2001introduction} problem at the arbitration phase (at most five line cards per CPU socket), enables us to cache the solution space and implement scheduling using constant-time lookups.

As shown in Figure~\ref{HyMoS-Arch}, HyMoS implements more than one datapath to leverage from highly flexible software-based packet processing in addition to P4 compatibility:
\begin{itemize}
    \item P4 datapath: P4-compatible NICs process packets and directly send them to egress NIC/s.
    \item P4 + DPDK datapath: NICs send packets to CPU for additional processing. 
    \item P4 + GPU datapath: Packets are sent to Graphical Processing Unit (GPU) for more complex applications such as deep packet inspection~\cite{sun2013fast,smith2009evaluating}.
\end{itemize}

Multiple datapaths are implemented by instantiating additional virtual interfaces on each NIC to serve as virtual queues for DPDK/GPU destinations.

\subsection{Smart NICs as Line Cards}

%\begin{figure}[b]
%    \centering
%    \includestandalone[width=0.42\textwidth]{figs/smartNIC}
%    \caption{Reconfigurable switch at smart NIC}
%    \label{typicalSmartNIC}
%\end{figure}

State-of-the-art smart NICs~\cite{nfp4000, ozdag2012intel, xpliant, zilberman2014netfpga} offer a reconfigurable hardware switch with multiple physical ports and tens of Single-Root I/O Virtualization (SR-IOV)~\cite{pci2010sriov} virtual interfaces.
These devices support processing packets at stunning rates beyond 100Gbps per direction.
%Figure~\ref{typicalSmartNIC} illustrates a typical smart NIC.
In addition to standard P4 actions, some even support defining custom packet processing actions~\cite{nfp4000}.

Utilizing smart NICs as line cards enables hardware-accelerated packet processing at line rate, P4-programmable datapath, and a modular design for end-users that allows them to choose line cards and customize switch port rate/count. 

\begin{figure}[t]
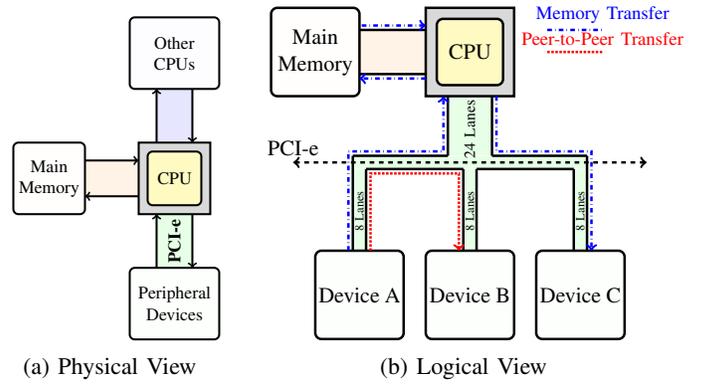

    \begin{subfigure}[t]{0.18\textwidth}
        \centering
        \includestandalone[width=\textwidth]{PCIe-1}
        \caption{Physical View}\label{PView}
    \end{subfigure}
    \begin{subfigure}[t]{0.33\textwidth}
        \centering
        \includestandalone[width=\textwidth]{PCIe-2}
        \caption{Logical View}\label{LView}
    \end{subfigure}
    \caption{PCI Express Architecture}
    \label{PCIe-views}
\end{figure}

\begin{table}[b]
    \centering
    {\small
    \begin{tabular}[] {c|cccccc}
        Link Width & x1 & x2 & x4 & x8 & x16 \\
        \hline
        Gen1 Bandwidth (GB/s) & 0.5 & 1 & 2 & 4 & 8 \\
        Gen2 Bandwidth (GB/s) &  1 & 2 & 4 & 8 & 16 \\
        Gen3 Bandwidth (GB/s) & $\sim$2 & $\sim$4 & $\sim$8 & $\sim$16 & $\sim$32 \\
    \end{tabular}
    \caption{PCI Express Evolution \cite{jackson2012pci}}
    \label{PCIe-Evolution}
    }
\end{table}

\subsection{PCI-e Interface as Switch Fabric}
PCI-e~\cite{pci2010rev3} is the interface that interconnects CPU and peripheral devices in the X86 architecture.
Modern implementations of PCI-e, as shown in Figure~\ref{PView}, directly attach peripheral devices to CPU with a PCI-e link which is a point-to-point dual simplex connection of up to 32 lanes.
Table~\ref{PCIe-Evolution} demonstrates how PCI-e standard has evolved over the years to provide more bandwidth for peripheral devices.
A third generation X8 link has enough bandwidth to support multiple 10/25/40G interfaces.
The fourth generation of PCI express, expected to be released this year, doubles third-generation bandwidth making 100G or dense 40G switches a possibility.

As illustrated in Figure~\ref{LView}, in addition to supporting memory operations, PCI-e supports peer-to-peer device transfers.
As opposed to the DPDK model in which packets traverse through the main memory for a transfer from device A to device C, a peer-to-peer PCI-e transaction transfers a packet from device A to device B directly without CPU/Memory involvement.
HyMoS relies on peer-to-peer transactions to switch packets between the cards.

SR-IOV allows instantiation of virtual interfaces on NIC cards.
Virtual interfaces have unique address and ingress/egress queues, and line cards implement a VOQ structure by instantiating such interfaces for each possible destination i.e. other line cards or CPU/GPU datapath.

\begin{figure}[t]
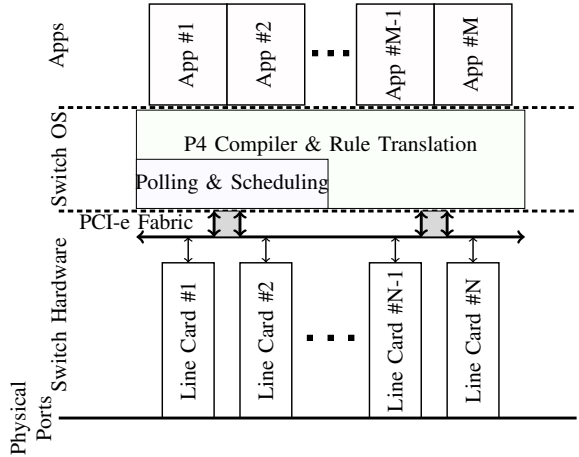

    \centering
    \includestandalone[width=0.425\textwidth]{modularity}
    \caption{Modular switch design.}
    \label{VSF}
\end{figure}

\subsection{Modularity in Software and Hardware}

HyMoS relies on CPU to implement a pipelined polling and scheduling process.
Depending on the choice of CPU users may have a number of spare cores.
As shown in Figure~\ref{VSF}, HyMoS brings in modularity in software by making an open platform and letting users install SDN agents.
For example, HyMoS users may install DIFANE~\cite{yu2010scalable}, NetPlumber~\cite{kazemian2013real}, and Flowvisor~\cite{sherwood2009flowvisor} agents for smart placement of rules on line cards, verification of policies, and network virtualization, respectively.
Cloud networking services can also be offloaded to HyMoS for faster deployment and robust control of cloud services.

We put switch operators in charge of hardware selection.
Number and type of CPU, the amount of memory, GPU packet processors as well as line cards can be customized.
The P4 datapath requires little to no CPU intervention as PCI-e peer-to-peer transactions do not consume CPU cycles.
To enable DPDK datapath or implement SDN agents, HyMoS should be configured with multiple cores at a high clock and sufficient memory.
GPU packet processors offer more complex and highly parallel network functions such as deep packet inspection at the cost of occupying some of the available PCI-e slots for line cards.

Switch operators also have a choice of line cards.
They can mix 10/25/40G cards on X8 PCI-e links.
A dual-port 40G card with breakout enables up to eight 10G ports enabling high port density.
Using 100G line cards as uplink ports is also possible, although these cards need an X16 or two X8 PCI-e links ideally connected to different CPU sockets.

%% file: MoPPetS-Arch.tex
\pagenumbering{gobble}

\begin{tikzpicture}

    \draw[<->] (1.8, 3.8) -- (8.05, 3.8) node[right] {PCI-e};
    \filldraw[fill=white, thick, white] (1.7, 3.7) rectangle (8.1, 3.9);
    \draw[ultra thick, <->] (1.8, 3.5) -- (8.05, 3.5) node[right] {Switch};
    \draw[<->] (1.8, 3.2) -- (8.05, 3.2) node[right] {Fabric};
    \filldraw[fill=white, thick, white] (1.7, 3.1) rectangle (8.1, 3.3);
    
    \draw[thick] (2.0, 0) rectangle(2.7, 2.8) node[pos=0.5, rotate=90] {Line Card \#1};  % CARD BORDER
    \draw[thick] (3.0, 0) rectangle(3.7, 2.8) node[pos=0.5, rotate=90] {Line Card \#2};  % CARD BORDER
    \draw[thick] (5.0, 0) rectangle(5.7, 2.8) node[pos=0.5, rotate=90] {Line Card \#n};  % CARD BORDER

    \foreach \i in {3.8, 4.2, 4.6}
    {
        \filldraw[fill=black] (\i, 1.45) rectangle(\i++0.1, 1.55);
    }

    \filldraw[ultra thick, fill=gray!30!white] (-0.8, 0.2) rectangle (1.8, 5.3);
    \filldraw[ultra thick, fill=yellow!30!white, rounded corners] (-0.6, 0.4) rectangle (1.6, 5.1) node[pos=0.7] {CPU\ \ \ \ \ \ \ };

    \filldraw[fill=white, thick, rounded corners] (-0.5, 4) rectangle(1.5, 5) node[pos=0.5] {Scheduler};
    \filldraw[fill=white, thick, rounded corners] (-0.5, 0.5) rectangle(1.5, 2.5) node[pos=0.5] {DPDK App};
    \draw[thick, rounded corners] (6, 0.5) rectangle(8, 2.5) node[pos=0.5] {GPU App};

    \draw[thick, <->] (1.6, 4.5) -- (2.35, 4.5) -- (2.35, 3.5);

    \draw[line width=0.4mm, blue, dashed, ->] (2.55, 2.8) -- (2.55, 3.2) -- (3.1, 3.2) -- (3.1, 2.8);
    \draw[line width=0.4mm, red, dotted, ->] (2.15, 2.8) -- (2.15, 3.2) -- (0.9, 3.2) -- (0.9, 2.5);
    \draw[line width=0.4mm, red, dotted, ->] (0.2, 2.5) -- (0.2, 3.35) -- (3.35, 3.35) -- (3.35, 2.8);
    \draw[line width=0.4mm, ForestGreen, dashdotted, ->] (5.35, 2.8) -- (5.35, 3.2) -- (6.8, 3.2) -- (6.8, 2.5);
    \draw[line width=0.4mm, ForestGreen, dashdotted, ->] (7.4, 2.5) -- (7.4, 3.4) -- (3.6, 3.4) -- (3.6, 2.8);

    \draw[thick, text=blue, blue, dashed] (4.3, 4.18) -- (5.5, 4.18) node[right] {P4 Datapath};
    \draw[thick, text=red, red, dotted] (4.3, 4.5) -- (5.5, 4.5) node[right] {P4 + DPDK Datapath};
    \draw[thick, text=ForestGreen, ForestGreen, dashdotted] (4.3, 4.82) -- (5.5, 4.82) node[right] {P4 + GPU Datapath};

\end{tikzpicture}

%% file: PCIe-1.tex
\pagenumbering{gobble}

\begin{tikzpicture}

    \filldraw[draw=white, fill=green!10!white] (1, 2) rectangle (2, 3.5);
    \filldraw[draw=white, fill=blue!10!white] (1, 5.5) rectangle (2, 7);
    \filldraw[draw=white, fill=orange!10!white] (0.5, 4) rectangle (-1, 5);

    \filldraw[ultra thick, rounded corners, fill=green!1!white] (0.25, 0) rectangle (2.75, 2) node[pos=0.5, text width=3cm] {\centering{\Large Peripheral\\Devices\\ }};
    \filldraw[ultra thick, fill=gray!30!white] (0.5, 3.5) rectangle (2.5, 5.5);
    \filldraw[ultra thick, fill=yellow!30!white, rounded corners] (0.75, 3.75) rectangle (2.25, 5.25) node[pos=0.5, text width=3cm] {\centering{\Large CPU\\}};
    \filldraw[ultra thick, rounded corners, fill=blue!1!white] (0.25, 7) rectangle (2.75, 9) node[pos=0.5, text width=3cm] {\centering{\Large Other\\CPUs\\ }};
    \filldraw[ultra thick, rounded corners, fill=orange!1!white] (-3, 3.5) rectangle (-1, 5.5) node[pos=0.5, text width=3cm] {\centering{\Large Main\\Memory\\ }};

    \node [rectangle, rotate=90] at(1.5,2.75) {\textbf{\Large PCI-e}};
    \draw [ultra thick, ->] (1, 2) -> (1, 3.5);
    \draw [ultra thick, <-] (2, 2) -> (2, 3.5);

    \draw [ultra thick, ->] (1, 5.5) -> (1, 7);
    \draw [ultra thick, <-] (2, 5.5) -> (2, 7);

    \draw [ultra thick, ->] (0.5, 4) -> (-1, 4);
    \draw [ultra thick, <-] (0.5, 5) -> (-1, 5);

\end{tikzpicture}

%% file: PCIe-2.tex
\pagenumbering{gobble}

\begin{tikzpicture}

    \filldraw[draw=white, fill=green!10!white] (1, 2) rectangle (2, 3.5);

    \filldraw[draw=white, fill=green!10!white] (-1.15, 1.85) rectangle (4.15, 2.15);

    \filldraw[draw=white, fill=green!10!white] (-1.15, 0) rectangle (-0.85, 1.85);
    \filldraw[draw=white, fill=green!10!white] ( 1.35, 0) rectangle ( 1.65, 1.85);
    \filldraw[draw=white, fill=green!10!white] ( 3.85, 0) rectangle ( 4.15, 1.85);

    \filldraw[draw=white, fill=orange!10!white] (0.5, 4) rectangle (-1, 5);

    \draw[ultra thick, <->, dashed] (-2.5, 2) node[above] {\Large PCI-e} -- (5.5, 2);

    \filldraw[ultra thick, rounded corners, fill=green!1!white] (-2, -2) rectangle (0, 0) node[pos=0.5, text width=3cm] {\centering{\Large Device A\\ }};
    \filldraw[ultra thick, rounded corners, fill=green!1!white] (0.5, -2) rectangle (2.5, 0) node[pos=0.5, text width=3cm] {\centering{\Large Device B\\ }};
    \filldraw[ultra thick, rounded corners, fill=green!1!white] (3, -2) rectangle (5, 0) node[pos=0.5, text width=3cm] {\centering{\Large Device C\\ }};
    \filldraw[ultra thick, fill=gray!30!white] (0.5, 3.5) rectangle (2.5, 5.5);
    \filldraw[ultra thick, fill=yellow!30!white, rounded corners] (0.75, 3.75) rectangle (2.25, 5.25) node[pos=0.5, text width=3cm] {\centering{\Large CPU\\}};
    \filldraw[ultra thick, rounded corners, fill=orange!1!white] (-3, 3.5) rectangle (-1, 5.5) node[pos=0.5, text width=3cm] {\centering{\Large Main\\Memory\\ }};

    \node [rectangle, rotate=90] at(1.5,2.75) {24 Lanes};
    \draw [ultra thick] (1, 2.15) -> (1, 3.5);
    \draw [ultra thick] (2, 2.15) -> (2, 3.5);

    \draw [ultra thick] (0.5, 4) -> (-1, 4);
    \draw [ultra thick] (0.5, 5) -> (-1, 5);

    \node [rectangle, rotate=90] at(1.5,0.9) {\footnotesize 8 Lanes};
    \draw [ultra thick] (1.35, 0) -- (1.35, 1.85);
    \draw [ultra thick] (1.65, 0) -- (1.65, 1.85);
    \draw [ultra thick] (-0.85, 1.85) -- (1.35, 1.85);
    \draw [ultra thick] (1.65, 1.85) -- (3.85, 1.85);

    \node [rectangle, rotate=90] at(-1,0.9) {\footnotesize 8 Lanes};
    \draw [ultra thick] (1, 2.15) -- (-1.15, 2.15);
    \draw [ultra thick] (-1.15, 0) -- (-1.15, 2.15);
    \draw [ultra thick] (-0.85, 0) -- (-0.85, 1.85);

    \node [rectangle, rotate=90] at(4,0.9) {\footnotesize 8 Lanes};
    \draw [ultra thick] (2, 2.15) -- (4.15, 2.15);
    \draw [ultra thick] (3.85, 0) -- (3.85, 1.85);
    \draw [ultra thick] (4.15, 0) -- (4.15, 2.15);

    \draw [densely dotted, red, ultra thick, ->] (-0.75, 0) -- (-0.75, 1.75) -- (1.25, 1.75) -- (1.25, 0);

    \draw [densely dotted, red, ultra thick] (3.25, 4.5) -- (4.5, 4.5) node[above] {\large Peer-to-Peer Transfer};
    \draw [dashdotted, blue, ultra thick] (3.25, 5) -- (4.5, 5) node[above] {\large Memory Transfer};

    \draw [dashdotted, blue, ultra thick, ->] (-1.25, 0) -- (-1.25, 2.25) -- (0.9, 2.25) -- (0.9, 3.5);
    \draw [dashdotted, blue, ultra thick, ->] (2.1, 3.5) -- (2.1, 2.25) -- (4.25, 2.25) -- (4.25, 0);
    \draw [dashdotted, blue, ultra thick, ->] (0.5, 3.9) -- (-1, 3.9);
    \draw [dashdotted, blue, ultra thick, <-] (0.5, 5.1) -- (-1, 5.1);

\end{tikzpicture}

%% file: modularity.tex
\pagenumbering{gobble}

\begin{tikzpicture}

    \newlength{\WW}
    \setlength{\WW}{10cm}

    \newlength{\HH}
    \setlength{\HH}{10cm}

    \draw[ultra thick] (0, 0) node[above, rotate=90, text width=0.15\WW] {\large \centering{Physical\\Ports\\ }} -- (\WW, 0);
    %--------------------------------------------------------- Line Cards
    \draw[thick] (0.20\WW, 0) rectangle(0.30\WW, 0.3\HH) node[pos=0.5, rotate=90] {\large Line Card \#1};  % CARD BORDER
    \draw[thick] (0.35\WW, 0) rectangle(0.45\WW, 0.3\HH) node[pos=0.5, rotate=90] {\large Line Card \#2};  % CARD BORDER
    \draw[thick] (0.60\WW, 0) rectangle(0.70\WW, 0.3\HH) node[pos=0.5, rotate=90] {\large Line Card \#N-1};  % CARD BORDER
    \draw[thick] (0.75\WW, 0) rectangle(0.85\WW, 0.3\HH) node[pos=0.5, rotate=90] {\large Line Card \#N};  % CARD BORDER

    \foreach \i in {0.48\WW, 0.52\WW, 0.56\WW}
    {
        \filldraw[fill=black] (\i, 0.15\HH) rectangle(\i++0.01\WW, 0.15\HH++0.01\WW);
    }

    %--------------------------------------------------------- PCI-e
    \draw[ultra thick, <->] (0.15\WW, 0.35\HH) node[above] {\large PCI-e Fabric} -- (0.9\WW, 0.35\HH);
    \foreach \i in {0.25\WW, 0.40\WW, 0.65\WW, 0.8\WW}
    {
        \draw[thick, <->] (\i, 0.3\HH) -- (\i, 0.35\HH); 
    }

    \node[rotate=90] at(0, 0.2\HH) {\large Switch Hardware};

    %--------------------------------------------------------- OS
    \draw[ultra thick, densely dashed] (0, 0.4\HH) -- (\WW, 0.4\HH);
    \filldraw[fill=green!2!white] (0.15\WW,0.405\HH) rectangle(0.9\WW,0.595\HH) node at(0.525\WW, 0.525\HH) {\large P4 Compiler \& Rule Translation};
    \filldraw[fill=blue!2!white] (0.15\WW,0.405\HH) rectangle(0.52\WW,0.50\HH) node[pos=0.5] {\large Polling \& Scheduling};

    \filldraw[fill=gray!30!white] (0.30\WW, 0.35\HH) rectangle(0.35\WW, 0.40\HH);
    \filldraw[fill=gray!30!white] (0.70\WW, 0.35\HH) rectangle(0.75\WW, 0.40\HH);

    \foreach \i in {0.30\WW, 0.35\WW, 0.70\WW, 0.75\WW}
    {
        \draw[ultra thick, <->] (\i, 0.35\HH) -- (\i, 0.40\HH); 
    }

    \node[rotate=90] at(0, 0.5\HH) {\large Switch OS};

    %--------------------------------------------------------- Software
    \draw[ultra thick, densely dashed] (0, 0.6\HH) -- (\WW, 0.6\HH);

    \draw[thick, fill=red!1!white] (0.175\WW, 0.605\WW) rectangle(0.325\WW, 0.8\HH) node[pos=0.5, rotate=90] {\large App \#1};  % CARD BORDER
    \draw[thick, fill=red!1!white] (0.325\WW, 0.605\WW) rectangle(0.475\WW, 0.8\HH) node[pos=0.5, rotate=90] {\large App \#2};  % CARD BORDER
    \draw[thick, fill=red!1!white] (0.575\WW, 0.605\WW) rectangle(0.725\WW, 0.8\HH) node[pos=0.5, rotate=90] {\large App \#M-1};  % CARD BORDER
    \draw[thick, fill=red!1!white] (0.725\WW, 0.605\WW) rectangle(0.875\WW, 0.8\HH) node[pos=0.5, rotate=90] {\large App \#M};  % CARD BORDER

    \foreach \i in {0.49\WW, 0.52\WW, 0.55\WW}
    {
        \filldraw[fill=black] (\i, 0.7\HH) rectangle(\i++0.01\WW, 0.7\HH++0.01\WW);
    }

    \node[rotate=90] at(0, 0.72\HH) {\large Apps};

\end{tikzpicture}

%% file: challenges.tex
\subsection{Challenges}\label{challenges}

\begin{figure}[t]
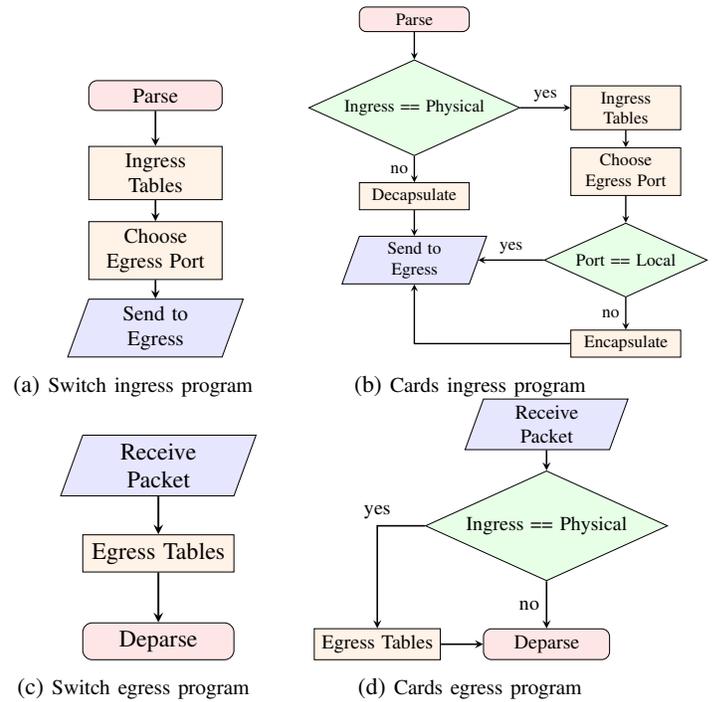

    \begin{subfigure}[t]{0.19\textwidth}
        \centering
        \includestandalone[width=0.85\textwidth]{P4S}
        \caption{\footnotesize Switch ingress program}
        \label{ingressSw}
    \end{subfigure}
    \begin{subfigure}[t]{0.29\textwidth}
        \centering
        \includestandalone[width=1.1\textwidth]{P4C}
        \caption{\footnotesize Cards ingress program}
        \label{ingressCard}
    \end{subfigure}

    \begin{subfigure}[b]{0.19\textwidth}
        \centering
        \includestandalone[width=0.95\textwidth]{P4SE}
        \caption{\footnotesize Switch egress program}
        \label{egressSw}
    \end{subfigure}
    \begin{subfigure}[b]{0.29\textwidth}
        \centering
        \includestandalone[width=\textwidth]{P4CE}
        \caption{\footnotesize Cards egress program}
        \label{egressCard}
    \end{subfigure}
    \caption{P4 Translation}
\end{figure}

\subsubsection{HyMoS' Compiler}\label{moppetsCompiler}

HyMoS relies on line cards to create a VOQ structure to implement internal packet switching between cards. 
As a result, input P4 program, $P$, which describes the behavior of the switch should be translated to $P_i$ P4 programs for individual line cards to implement table lookups for internal packet switching.
Therefore, we design a P4 compiler for HyMoS to add additional tables and derive $P_i$, line card programs, from $P$, the program for the switch. 

The implementation of HyMoS compiler is relatively straight forward.
Figures~\ref{ingressSw} and~\ref{egressSw} show the switch program for ingress and egress pipelines respectively.
As shown in Figure~\ref{ingressCard}, HyMoS compiler adds two additional tables at the beginning and end of ingress control flow to perform internal port lookups.
Packets destined to other line cards should be labeled with their destination port numbers since egress port specified in the P4 code is a metadata local to line card, and it is not transferred to the egress line card. 
MAC-in-MAC encapsulation is utilized to transfer destination port number along with the packet.
Added Ethernet header uses ingress port number as source MAC  and egress port number as destination MAC.
HyMoS encapsulation uses an unused Ethernet type for the parser to detect encapsulated packets.

The table at the end of the ingress pipeline decides whether the output port is local to current NIC or not.
If the destination port is local, then the packet will be sent to local egress port without any additional processing.
In cases where destination port is on another NIC, this table encapsulates the packet and queues it on the dedicated virtual interface for the egress line card.

As shown in Figure~\ref{egressCard}, egress table operations are also performed at ingress line card.
The table at the beginning of egress control flow matches on the input port.
Packets received on physical ports are processed as specified by $P$, while packets received on virtual ports are merely sent to their specified egress port.

In addition to adding tables mentioned above, the parser should also be modified to support MAC-in-MAC encapsulation and extract ingress/egress ports as metadata for encapsulated packets.

\begin{figure}
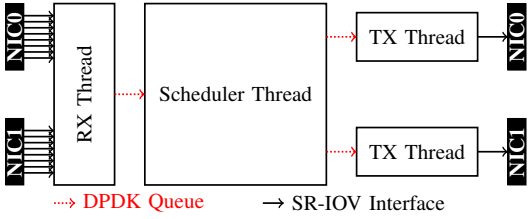

    \centering
    \includestandalone[width=0.395\textwidth]{scheduler}
    \caption{DPDK pipeline of HyMoS scheduler}
    \label{moppetsSchedule}
\end{figure}

\subsubsection{DPDK Packet Scheduler}

HyMoS DPDK scheduler implements IEEE802.1p~\cite{ek1999ieee}, which defines 8 priority classes and provides QoS at L2.
Supporting this protocol is a must for NFV targets, because it can be used as a building block to implement more advanced QoS protocols and/or guarantee SLAs.
However, the current generation of P4 language and P4-compatible devices are unable to support this protocol due to language limitations.

Our implementation relies on NIC cards to parse 802.1p priority class (which is part of the VLAN header) and enqueue processed packets in different SR-IOV virtual interfaces according to their priority class.
As mentioned before, HyMoS takes advantage of SR-IOV interfaces to virtualize VOQ structure.
With added support for 802.1p protocol, NIC cards create 8 queues per destination. In other words, on a switch with $N$ line cards, each line card creates $8(N-1)$ virtual queues.
Figure~\ref{moppetsSchedule} illustrates a simple example of this structure with 2 NICs.
In HyMoS' pipelined scheduler, a receiver thread polls SR-IOV interfaces and updates 8 demand matrices (one per priority class) for next-stage scheduler.
The scheduler thread greedily solves the problem for each priority class.
It starts with highest priority and looks up the solution to a bi-partite matching problem with maximum weight objective corresponding to current priority class, marks granted queues and iterates to the next priority level.
At the last stage of the pipeline one transmitter thread per line card transfers packets from marked queues to their destination.
Results in Section~\ref{evaluation} show that using single-producer multi-consumer queues from DPDK standard library and 4 CPU threads this architecture can transfer more than $100Gbps$ of traffic between HyMoS line cards.

\begin{figure}[t]
    \centering
    \includestandalone[width=0.395\textwidth]{testbedTopology}
    \caption{Testbed topology.}
    \label{testbedTopology}
\end{figure}

\begin{figure*}[t]
    \begin{subfigure}[t]{0.32\textwidth}
        \centering
        \includegraphics[width=\textwidth]{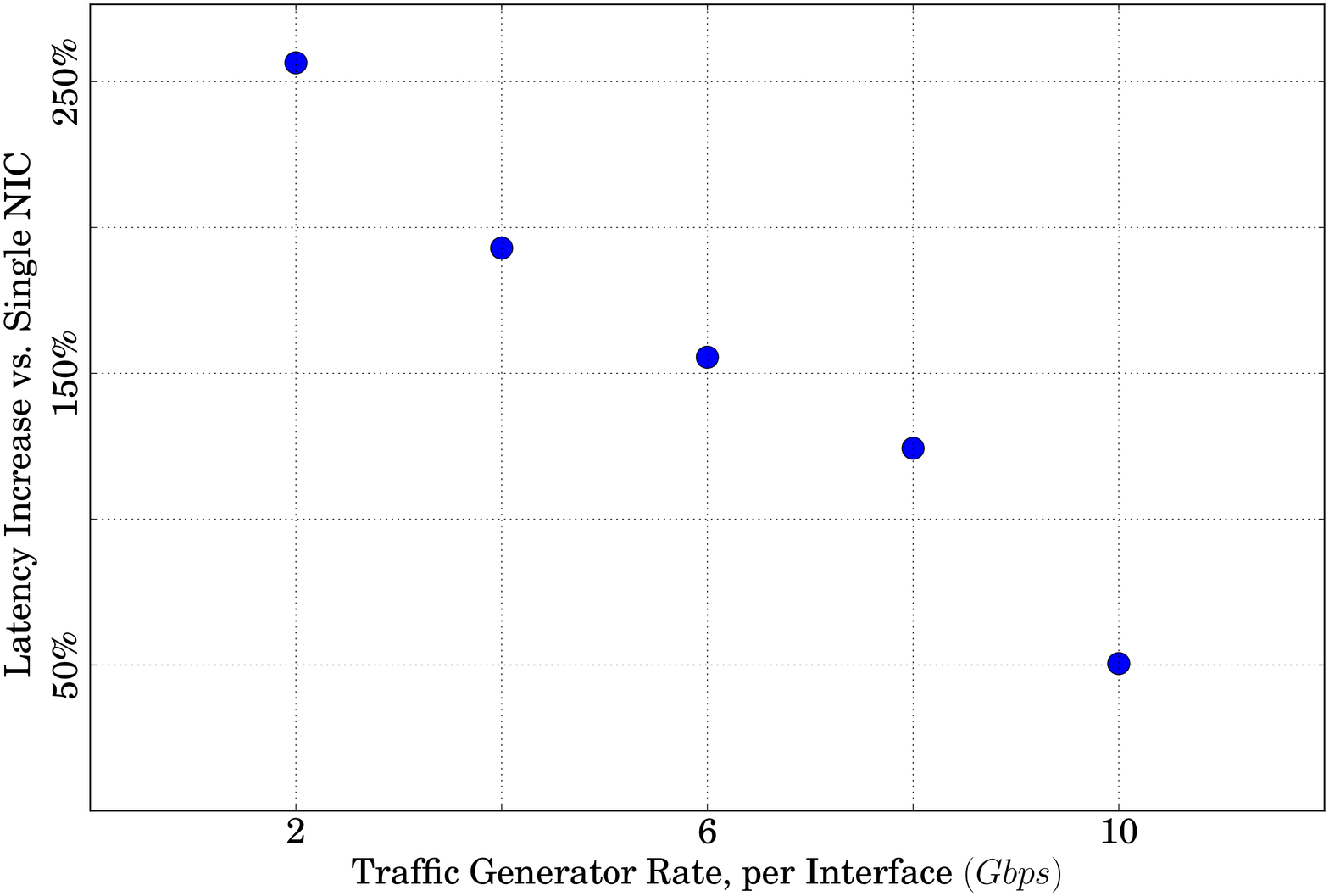}
        \caption{Normalized latency vs. switch load.}
        \label{moppets-load}
    \end{subfigure}
    \begin{subfigure}[t]{0.32\textwidth}
        \centering
        \includegraphics[width=\textwidth]{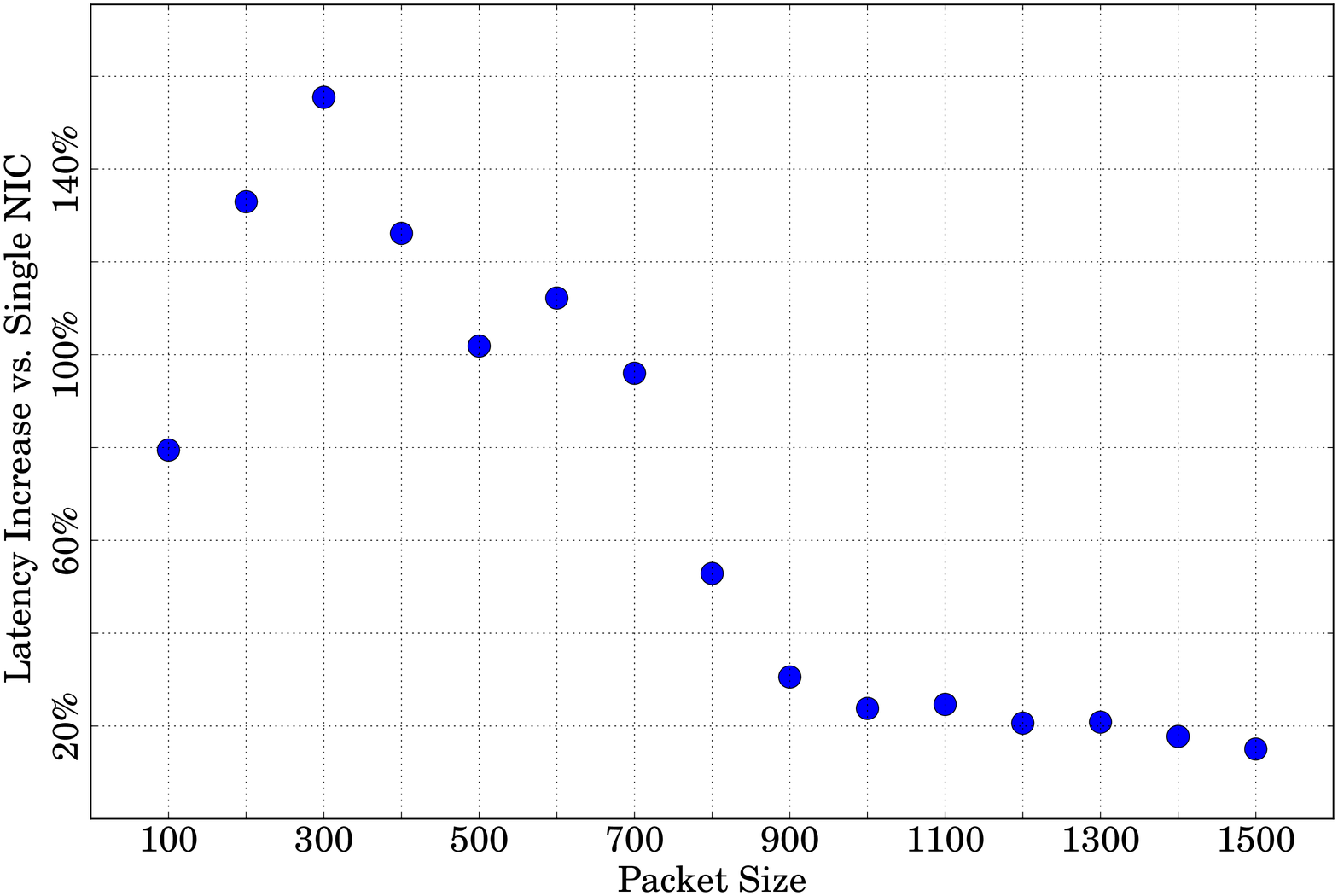}
        \caption{Normalized latency vs. packet size.}
        \label{moppets-size}
    \end{subfigure}
    \begin{subfigure}[t]{0.32\textwidth}
        \centering
        \includegraphics[width=\textwidth]{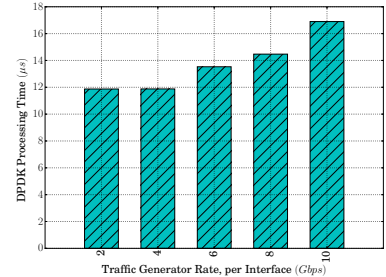}
        \caption{DPDK processing time vs. switch load.}
        \label{dpdk-load}
    \end{subfigure}
    \caption{HyMoS' relative performance.}
\end{figure*}

%% file: P4S.tex
\pagenumbering{gobble}

\tikzstyle{startstop} = [rectangle, rounded corners, minimum width=2cm, text width=2cm, minimum height=0.5cm,text centered, draw=black, fill=red!10]
\tikzstyle{io} = [trapezium, trapezium left angle=70, trapezium right angle=110, minimum width=2cm, text width=2cm, minimum height=0.5cm, text centered, draw=black, fill=blue!10]
\tikzstyle{process} = [rectangle, minimum width=2cm, text width=2cm, minimum height=0.5cm, text centered, draw=black, fill=orange!10]
\tikzstyle{decision} = [diamond, minimum width=2cm, text width=2cm, minimum height=0.5cm, text centered, draw=black, fill=green!10]
\tikzstyle{arrow} = [thick,->,>=stealth]

\begin{tikzpicture}[node distance=1.3cm]
    \node[startstop] at(0,0) (parse) {Parse};
    \node[process, below of=parse] (tables) {Ingress Tables};
    \node[process, below of=tables] (port) {Choose Egress Port};
    \node[io, below of=port] (send) {Send to Egress};
    \draw[arrow] (parse) -- (tables);
    \draw[arrow] (tables) -- (port);
    \draw[arrow] (port) -- (send);
\end{tikzpicture}

%% file: P4C.tex
\pagenumbering{gobble}

\tikzstyle{startstop} = [rectangle, rounded corners, minimum width=2cm, text width=2cm, minimum height=0.5cm,text centered, draw=black, fill=red!10]
\tikzstyle{io} = [trapezium, trapezium left angle=70, trapezium right angle=110, minimum width=2cm, text width=2cm, minimum height=0.5cm, text centered, draw=black, fill=blue!10]
\tikzstyle{process} = [rectangle, minimum width=2cm, text width=2cm, minimum height=0.5cm, text centered, draw=black, fill=orange!10]
\tikzstyle{decision} = [diamond, aspect=2.2, text centered, draw=black, fill=green!10]
\tikzstyle{arrow} = [thick,->,>=stealth]

\begin{tikzpicture}[node distance=1.3cm]
    \node[startstop] at(0,0) (parse) {Parse};
    \node[decision, below of=parse, yshift=-0.5cm] (phyD) {Ingress == Physical};
    \node[process, right of=phyD, xshift=3cm] (tables) {Ingress Tables};
    \node[process, below of=tables] (port) {Choose Egress Port};
    \node[decision, below of=port, yshift=-0.5cm] (locD) {Port == Local};
    \node[process, below of=locD, yshift=-0.4cm] (add) {Encapsulate};

    \node[process, below of=phyD, yshift=-0.5cm] (remove) {Decapsulate};
    \node[io, left of=locD, xshift=-3cm] (sendl) {Send to Egress};

    \draw[arrow] (parse) -- (phyD);
    \draw[arrow] (phyD) -- node[anchor=east] {no} (remove);
    \draw[arrow] (remove) -- (sendl);
    \draw[arrow] (phyD) -- node[anchor=south] {yes} (tables);
    \draw[arrow] (tables) -- (port);
    \draw[arrow] (port) -- (locD);
    \draw[arrow] (locD) -- node [anchor=east] {no} (add);
    \draw[arrow] (add) -| (sendl);
    \draw[arrow] (locD) -- node[anchor=south] {yes} (sendl);
\end{tikzpicture}

%% file: P4SE.tex
\pagenumbering{gobble}

\tikzstyle{startstop} = [rectangle, rounded corners, minimum width=2cm, text width=2cm, minimum height=0.5cm,text centered, draw=black, fill=red!10]
\tikzstyle{io} = [trapezium, trapezium left angle=70, trapezium right angle=110, minimum width=2cm, text width=2cm, minimum height=0.5cm, text centered, draw=black, fill=blue!10]
\tikzstyle{process} = [rectangle, minimum width=2cm, text width=2cm, minimum height=0.5cm, text centered, draw=black, fill=orange!10]
\tikzstyle{decision} = [diamond, minimum width=2cm, text width=2cm, minimum height=0.5cm, text centered, draw=black, fill=green!10]
\tikzstyle{arrow} = [thick,->,>=stealth]

\begin{tikzpicture}[node distance=1.3cm]
    \node[io] at(0,0) (parse) {Receive Packet};
    \node[process, below of=parse] (tables) {Egress Tables};
    \node[startstop, below of=tables] (deparse) {Deparse};
    \draw[arrow] (parse) -- (tables);
    \draw[arrow] (tables) -- (deparse);
\end{tikzpicture}

%% file: P4CE.tex
\pagenumbering{gobble}

\tikzstyle{startstop} = [rectangle, rounded corners, minimum width=2cm, text width=2cm, minimum height=0.5cm,text centered, draw=black, fill=red!10]
\tikzstyle{io} = [trapezium, trapezium left angle=70, trapezium right angle=110, minimum width=2cm, text width=2cm, minimum height=0.5cm, text centered, draw=black, fill=blue!10]
\tikzstyle{process} = [rectangle, minimum width=2cm, text width=2cm, minimum height=0.5cm, text centered, draw=black, fill=orange!10]
\tikzstyle{decision} = [diamond, aspect=2.2, text centered, draw=black, fill=green!10]
\tikzstyle{arrow} = [thick,->,>=stealth]

\begin{tikzpicture}[node distance=1.3cm]
    \node[io] at(0,0) (parse) {Receive Packet};
    \node[decision, below of=parse, yshift=-0.5cm] (dec) {Ingress == Physical};
    \node[process, below of=dec, yshift=-0.8cm, xshift=-3cm] (tables) {Egress Tables};
    \node[startstop, below of=dec, yshift=-0.8cm] (deparse) {Deparse};
    \draw[arrow] (parse) -- (dec);
    \draw[arrow] (dec) -| node[anchor=south] {yes} (tables);
    \draw[arrow] (tables) -- (deparse);
    \draw[arrow] (dec) -- node[anchor=east] {no} (deparse);
\end{tikzpicture}

%% file: scheduler.tex
\pagenumbering{gobble}

\begin{tikzpicture}
    
    \draw[thick] (1.5, 0) rectangle(4.5, 3) node[pos=0.5] {Scheduler Thread};

    \draw[thick] (0, 0) rectangle(1, 3) node[pos=0.5, rotate=90] {RX Thread};

    \draw[thick] (5, 0.15) rectangle(7, 0.95) node[pos=0.5] {TX Thread};
    \draw[thick] (5, 2.05) rectangle(7, 2.85) node[pos=0.5] {TX Thread};
    
    \draw[thick, densely dotted, red, ->] (1, 1.5) -- (1.5, 1.5);
    \draw[thick, densely dotted, red, ->] (4.5, 0.55) -- (5, 0.55);
    \draw[thick, densely dotted, red, ->] (4.5, 2.45) -- (5, 2.45);
    
    \draw[thick, ->] (-0.5, 0.2) -- (0, 0.2);
    \draw[thick, ->] (-0.5, 0.3) -- (0, 0.3);
    \draw[thick, ->] (-0.5, 0.4) -- (0, 0.4);
    \draw[thick, ->] (-0.5, 0.5) -- (0, 0.5);
    \draw[thick, ->] (-0.5, 0.6) -- (0, 0.6);
    \draw[thick, ->] (-0.5, 0.7) -- (0, 0.7);
    \draw[thick, ->] (-0.5, 0.8) -- (0, 0.8);
    \draw[thick, ->] (-0.5, 0.9) -- (0, 0.9);
    \draw[thick, ->] (-0.5, 2.1) -- (0, 2.1);
    \draw[thick, ->] (-0.5, 2.2) -- (0, 2.2);
    \draw[thick, ->] (-0.5, 2.3) -- (0, 2.3);
    \draw[thick, ->] (-0.5, 2.4) -- (0, 2.4);
    \draw[thick, ->] (-0.5, 2.5) -- (0, 2.5);
    \draw[thick, ->] (-0.5, 2.6) -- (0, 2.6);
    \draw[thick, ->] (-0.5, 2.7) -- (0, 2.7);
    \draw[thick, ->] (-0.5, 2.8) -- (0, 2.8);
    \draw[thick, ->] (7, 0.55) -- (7.5, 0.55);
    \draw[thick, ->] (7, 2.45) -- (7.5, 2.45);

    \draw[thick, ->] (3.45, -0.3) -- (3.80, -0.3) node[right] {SR-IOV Interface};
    \draw[thick, densely dotted, red, ->] (0, -0.3) -- (0.35, -0.3) node[right] {DPDK Queue};

    \filldraw[fill=black, text=white] (-0.8, 1.9) rectangle(-0.5, 3.0) node[pos=0.5, rotate=90] {\bf NIC0}; % Ports
    \filldraw[fill=black, text=white] (-0.8, 0) rectangle(-0.5, 1.1) node[pos=0.5, rotate=90] {\bf NIC1}; % Ports

    \filldraw[fill=black, text=white] (7.5, 1.9) rectangle(7.8, 3.0) node[pos=0.5, rotate=90] {\bf NIC0}; % Ports
    \filldraw[fill=black, text=white] (7.5, 0) rectangle(7.8, 1.1) node[pos=0.5, rotate=90] {\bf NIC1}; % Ports

    %\draw[thick] (8, 0) rectangle(9, 5) node[pos=0.5, rotate=90] {MoonGen Receiver};

    %\draw[thick] (3.025, 3.5) rectangle(5.975, 4.5) node[pos=0.5] {Line Card};
    %\draw[thick] (3.025, 0.5) rectangle(5.975, 1.5) node[pos=0.5] {Line Card};

    %\draw[line width=0.4mm, ->] (1, 3.7) -- (3, 3.7);
    %\draw[line width=0.4mm, ->] (1, 3.9) -- (3, 3.9);
    %\draw[line width=0.4mm, ->] (1, 4.1) -- (3, 4.1);
    %\draw[line width=0.4mm, ->] (1, 4.3) -- (3, 4.3);

    %\draw[line width=0.4mm, ->] (6, 3.7) -- (8, 3.7);
    %\draw[line width=0.4mm, ->] (6, 3.9) -- (8, 3.9);
    %\draw[line width=0.4mm, ->] (6, 4.1) -- (8, 4.1);
    %\draw[line width=0.4mm, ->] (6, 4.3) -- (8, 4.3);

    %\draw[line width=0.4mm, ->] (1, 0.7) -- (3, 0.7);
    %\draw[line width=0.4mm, ->] (1, 0.9) -- (3, 0.9);
    %\draw[line width=0.4mm, ->] (1, 1.1) -- (3, 1.1);
    %\draw[line width=0.4mm, ->] (1, 1.3) -- (3, 1.3);

    %\draw[line width=0.4mm, ->] (6, 0.7) -- (8, 0.7);
    %\draw[line width=0.4mm, ->] (6, 0.9) -- (8, 0.9);
    %\draw[line width=0.4mm, ->] (6, 1.1) -- (8, 1.1);
    %\draw[line width=0.4mm, ->] (6, 1.3) -- (8, 1.3);

    %\draw[thick, rounded corners, dashed, red] (0.9, 3.4) rectangle(8.1, 4.6);
    %\draw[thick, rounded corners, dashdotted, blue] (0.8, 0.3) rectangle(8.2, 4.7);

    %\draw[thick, ->] (3.45, -0.3) -- (3.80, -0.3) node[right] {10G Link};
    %\draw[thick, dashed, red] (0, -0.3) -- (0.35, -0.3) node[right] {Baseline Testbed};
    %\draw[thick, dashdotted, blue] (5.75, -0.3) -- (6.1, -0.3) node[right] {MoPPetS Testbed};

\end{tikzpicture}

%% file: testbedTopology.tex
\pagenumbering{gobble}

\begin{tikzpicture}
    
    \draw[thick] (3, 0) rectangle(6, 5) node[pos=0.5] {HyMoS};

    \draw[thick] (0, 0) rectangle(1, 5) node[pos=0.5, rotate=90] {MoonGen Transmitter};
    \draw[thick] (8, 0) rectangle(9, 5) node[pos=0.5, rotate=90] {MoonGen Receiver};

    \draw[thick] (3.025, 3.5) rectangle(5.975, 4.5) node[pos=0.5] {Line Card};
    \draw[thick] (3.025, 0.5) rectangle(5.975, 1.5) node[pos=0.5] {Line Card};

    \draw[line width=0.4mm, ->] (1, 3.7) -- (3, 3.7);
    \draw[line width=0.4mm, ->] (1, 3.9) -- (3, 3.9);
    \draw[line width=0.4mm, ->] (1, 4.1) -- (3, 4.1);
    \draw[line width=0.4mm, ->] (1, 4.3) -- (3, 4.3);

    \draw[line width=0.4mm, ->] (6, 3.7) -- (8, 3.7);
    \draw[line width=0.4mm, ->] (6, 3.9) -- (8, 3.9);
    \draw[line width=0.4mm, ->] (6, 4.1) -- (8, 4.1);
    \draw[line width=0.4mm, ->] (6, 4.3) -- (8, 4.3);

    \draw[line width=0.4mm, ->] (1, 0.7) -- (3, 0.7);
    \draw[line width=0.4mm, ->] (1, 0.9) -- (3, 0.9);
    \draw[line width=0.4mm, ->] (1, 1.1) -- (3, 1.1);
    \draw[line width=0.4mm, ->] (1, 1.3) -- (3, 1.3);

    \draw[line width=0.4mm, ->] (6, 0.7) -- (8, 0.7);
    \draw[line width=0.4mm, ->] (6, 0.9) -- (8, 0.9);
    \draw[line width=0.4mm, ->] (6, 1.1) -- (8, 1.1);
    \draw[line width=0.4mm, ->] (6, 1.3) -- (8, 1.3);

    \draw[thick, rounded corners, dashed, red] (0.9, 3.4) rectangle(8.1, 4.6);
    \draw[thick, rounded corners, dashdotted, blue] (0.8, 0.3) rectangle(8.2, 4.7);

    \draw[thick, ->] (3.45, -0.3) -- (3.80, -0.3) node[right] {10G Link};
    \draw[thick, dashed, red] (0, -0.3) -- (0.35, -0.3) node[right] {Baseline Testbed};
    \draw[thick, dashdotted, blue] (5.75, -0.3) -- (6.1, -0.3) node[right] {HyMoS Testbed};

\end{tikzpicture}

%% file: evaluation.tex
\section{Preliminary Results} \label{evaluation}

%We use a single P4-compatible NIC to create a baseline for evaluating HyMoS performance.
%Processing time at the switch is used as the main metric for evaluating packet forwarding performance.

Our HyMoS prototype implements P4+DPDK datapath.
Figure~\ref{testbedTopology} shows the topology of HyMoS testbed.
Our early implementation uses two dual-port 40G NFP6000~\cite{nfp4000} smart NICs on a dual-socket server with Intel Xeon E5 2690 V3 twelve-core 2.6GHz CPUs.
NICs are attached to the same Non-uniform Memory Access~\cite{lameter2013numa} (NUMA) node on the server and HyMoS' DPDK scheduler is implemented using 4 cores from the same NUMA node.
The second CPU is not used in this experiment.
40G ports on the NICs are configured to work as quad 10G interfaces allowing us to implement a 16-port 10G HyMoS.

MoonGen~\cite{emmerich2015moongen} is utilized to generate multiple 10G streams with variable packet sizes.
A traffic generator that uses four quad-port Intel X710 NICs is connected to to HyMoS line cards.

To create a baseline for performance, we have installed an L3 router P4 program on an NFP-6000 NIC.
The P4 code implements IP Longest Prefix Matching (LPM), VLAN tagging, and Ethernet Forwarding Information Base (FIB) tables.
As shown in Figure~\ref{testbedTopology}, four 10G interfaces send UDP traffic with uniformly distributed random source and destination IP addresses to 4 ports of the smart NIC.
Rules are installed on the NIC card to put the remaining four interfaces on four different IP subnets which are connected to MoonGen receivers.
We defined end-to-end delay - which approximates the processing time at switch - as the performance metric for this experiment.
MoonGen is configured to measure end-to-end delay using hardware timestamps.

HyMoS is evaluated under similar settings.
The same L3 router P4 code is translated to HyMoS using method discussed in \ref{moppetsCompiler}.
As shown in Figure~\ref{testbedTopology}, eight traffic generator interfaces are sending uniform traffic to HyMoS (four interfaces on each line card) and the remaining eight interfaces receive the traffic.
Similar to the baseline, end-to-end delay is measured using hardware timestamps at traffic generator NICs.

Figure~\ref{moppets-load} presents the average increase in switch processing time of HyMoS compared to baseline NFP6000 under variable load.
In this scenario, the traffic generator sends 800 Byte packets at specified rates varying between 2Gbps to 10Gbps per interface for 5 minutes in each measurement.
Figure~\ref{moppets-size} compares the increase in processing time of HyMoS with variable packet sizes.
In this case, packets with specified lengths are generated at line rate for 5 minutes for each measurement. 
Our measurements show that HyMoS' performance comes very close to baseline (less than 20\% penalty in processing time) at line rate with large packets, which is very promising given that this implementation of HyMoS offers more flexible scheduling, stateful packet processing at DPDK, and twice the ports of a single NIC.
At slower rates and with small packets, however, performance penalty could be larger.

As a micro-benchmark, DPDK transfer time between the two line cards is measured by installing receive and transmit callbacks.
This quantity is equal to the queuing delay in addition to the processing delay incurred by HyMoS DPDK scheduler.
Figure~\ref{dpdk-load} present the average transfer times under variable load with 800 Byte packets and transfer times under variable load.
Similar to previous results, the average values are taken from 5-minute long measurements.
Based on these figures, we conclude that DPDK processing time outweighs NFP6000 processing time at slower rates and for smaller packets.
However, in processing large packets at line rate processing time at NIC outweighs DPDK processing time which effectively makes the cost of additional HyMoS features negligible.

%% file: related.tex
\section{Related Works}\label{related}
HyMoS is built on top of four pillars.
\textbf{P4}~\cite{bosshart2014p4}, a universal programming language to describe the behavior of packet processors.
\textbf{DPDK}~\cite{intel2014data}, a set of libraries that enable programmable packet processing at software.
And recent advances in X86 computing, \textbf{Smart NICs}\cite{nfp4000,zilberman2014netfpga,ozdag2012intel,xpliant} that makes high-throughput packet processing at hardware possible and \textbf{PCI Express interface}~\cite{pci2010rev3}, a high-speed interconnect for the peripheral devices.
Donard~\cite{donard} and Direct-GPU~\cite{directgpu} are two recent technologies that use smart NICs and PCI-e peer-to-peer transactions to realize remote direct memory access (RDMA) for SSD/NVMe storage and GPU memory respectively.

ServerSwitch~\cite{lu2011serverswitch} is one of the earliest works that utilized PCI-e as a cost-effective alternative for Ethernet switching.
Belonging to pre-SDN era, ServerSwitch does not offer much programmability atop Ethernet.
As a programmable switch, HyMoS is closely related to PISA~\cite{bosshart2013forwarding}, PISCES~\cite{shahbaz2016pisces}, and OCP Wedge switch~\cite{ocp2015wedge}.
\cite{bosshart2013forwarding} processes packets using a domain-specific hardware, whereas \cite{bosshart2013forwarding,shahbaz2016pisces} rely on software to make a programmable data path.
ClickNP~\cite{li2016clicknp} and NetVM~\cite{hwang2015netvm} are other examples of highly programmable devices designed for NFV without supporting P4 DSL.
ClickNP offers FPGA-based hardware acceleration achieving low-latency and high throughput for packet forwarding, however, it comes at the cost of complicated design and deployment of new network functions due to not supporting a DSL similar to P4.
NetVM solely relies on DPDK for packet processing offering a highly programmable solution with X86 performance limitations.
Unlike existing solutions, we take the middle ground and leverage both, hardware and software, to process the packets

HyMoS is a modular platform that enables the implementation of recent advances in NFV/SDN including but not limited to rule caching and management~\cite{yu2010scalable,katta2014infinite,yan2014cab}, network function virtualization~\cite{wood2015toward,price2012opnfv}, control plane virtualization layers~\cite{hancock2016hyper4,sherwood2009flowvisor}, rule verification~\cite{kazemian2013real,foster2011frenetic,khurshid2012veriflow}, and flexible/efficient data plane design\cite{intel2014data,hwang2015netvm,cerrato2014supporting}. 

HyMoS also builds on top of extensive research in switch design, most notably, scheduling in virtual output queued switches\cite{mckeown1999achieving,mckeown1999islip,karol1987input} and its relation to the classic bipartite matching problem in graphs~\cite{west2001introduction}.

Whippersnapper~\cite{dang2017whippersnapper} is a framework for benchmarking P4-compatible devices which we plan to use in the future for a more comprehensive comparison of our design to existing solutions. 

%% file: future.tex
\section{Conclusion} \label{future}
HyMoS is a programmable modular switch designed for NFV applications at networks' edge.
Its hybrid hardware/software design based on commodity servers enables modularity and a high degree of programmability. 
Unlike solutions that rely only on hardware or on software, HyMoS brings in the best of both worlds by utilizing P4-compatible NICs and PCI-e backplane to enable flexible packet processing and switching at line rate.

In addition to supporting a hardware-accelerated P4-compatible datapath suited for virtualizing L2/L3 functions, HyMoS offers a P4+DPDK and P4+GPU datapaths geared towards advanced L4 and up NFV applications.
HyMoS improves the usability of P4 language in NFV applications by adding a programmable scheduler and enabling support for DPDK packet processing on top of P4.
Using a small testbed we show that HyMoS extra features will add a performance penalty, which is negligible especially at line rate and for large packets.

    %\item Optimization of HyMoS P4 compiler.
    %    While copying P4 program at line cards guarantees the as described operation of the switch, it may not be the smallest program.
    %    In other words, a more intelligent compiler may derive different P4 codes at line cards that are smaller in size compared to the switch program while implementing the same program in combination.

%% file: ms.bbl
\begin{thebibliography}{10}

\bibitem{mckeown2008openflow}
N.~McKeown, T.~Anderson, H.~Balakrishnan, G.~Parulkar, L.~Peterson, J.~Rexford,
  S.~Shenker, and J.~Turner, ``Openflow: enabling innovation in campus
  networks,'' {\em ACM SIGCOMM Computer Communication Review}, vol.~38, no.~2,
  pp.~69--74, 2008.

\bibitem{gude2008nox}
N.~Gude, T.~Koponen, J.~Pettit, B.~Pfaff, M.~Casado, N.~McKeown, and
  S.~Shenker, ``Nox: towards an operating system for networks,'' {\em ACM
  SIGCOMM Computer Communication Review}, vol.~38, no.~3, pp.~105--110, 2008.

\bibitem{yu2010scalable}
M.~Yu, J.~Rexford, M.~J. Freedman, and J.~Wang, ``Scalable flow-based
  networking with difane,'' {\em ACM SIGCOMM Computer Communication Review},
  vol.~40, no.~4, pp.~351--362, 2010.

\bibitem{katta2014infinite}
N.~Katta, O.~Alipourfard, J.~Rexford, and D.~Walker, ``Infinite cacheflow in
  software-defined networks,'' in {\em Proceedings of the third workshop on Hot
  topics in software defined networking}, pp.~175--180, ACM, 2014.

\bibitem{yan2014cab}
B.~Yan, Y.~Xu, H.~Xing, K.~Xi, and H.~J. Chao, ``Cab: A reactive wildcard rule
  caching system for software-defined networks,'' in {\em Proceedings of the
  third workshop on Hot topics in software defined networking}, pp.~163--168,
  ACM, 2014.

\bibitem{kim2013improving}
H.~Kim and N.~Feamster, ``Improving network management with software defined
  networking,'' {\em IEEE Communications Magazine}, vol.~51, no.~2,
  pp.~114--119, 2013.

\bibitem{porras2012security}
P.~Porras, S.~Shin, V.~Yegneswaran, M.~Fong, M.~Tyson, and G.~Gu, ``A security
  enforcement kernel for openflow networks,'' in {\em Proceedings of the first
  workshop on Hot topics in software defined networks}, pp.~121--126, ACM,
  2012.

\bibitem{kazemian2013real}
P.~Kazemian, M.~Chan, H.~Zeng, G.~Varghese, N.~McKeown, and S.~Whyte, ``Real
  time network policy checking using header space analysis.,'' in {\em NSDI},
  pp.~99--111, 2013.

\bibitem{khurshid2012veriflow}
A.~Khurshid, W.~Zhou, M.~Caesar, and P.~Godfrey, ``Veriflow: Verifying
  network-wide invariants in real time,'' {\em ACM SIGCOMM Computer
  Communication Review}, vol.~42, no.~4, pp.~467--472, 2012.

\bibitem{foster2011frenetic}
N.~Foster, R.~Harrison, M.~J. Freedman, C.~Monsanto, J.~Rexford, A.~Story, and
  D.~Walker, ``Frenetic: A network programming language,'' in {\em ACM Sigplan
  Notices}, vol.~46, pp.~279--291, ACM, 2011.

\bibitem{reich2013modular}
J.~Reich, C.~Monsanto, N.~Foster, J.~Rexford, and D.~Walker, ``Modular sdn
  programming with pyretic,'' {\em Technical Reprot of USENIX}, 2013.

\bibitem{yu2013software}
M.~Yu, L.~Jose, and R.~Miao, ``Software defined traffic measurement with
  opensketch.,'' in {\em NSDI}, vol.~13, pp.~29--42, 2013.

\bibitem{moshref2014dream}
M.~Moshref, M.~Yu, R.~Govindan, and A.~Vahdat, ``Dream: dynamic resource
  allocation for software-defined measurement,'' in {\em ACM SIGCOMM Computer
  Communication Review}, vol.~44, pp.~419--430, ACM, 2014.

\bibitem{bosshart2013forwarding}
P.~Bosshart, G.~Gibb, H.-S. Kim, G.~Varghese, N.~McKeown, M.~Izzard, F.~Mujica,
  and M.~Horowitz, ``Forwarding metamorphosis: Fast programmable match-action
  processing in hardware for sdn,'' in {\em ACM SIGCOMM Computer Communication
  Review}, vol.~43, pp.~99--110, ACM, 2013.

\bibitem{van2000domain}
A.~Van~Deursen, P.~Klint, J.~Visser, {\em et~al.}, ``Domain-specific languages:
  An annotated bibliography.,'' {\em Sigplan Notices}, vol.~35, no.~6,
  pp.~26--36, 2000.

\bibitem{bosshart2014p4}
P.~Bosshart, D.~Daly, G.~Gibb, M.~Izzard, N.~McKeown, J.~Rexford,
  C.~Schlesinger, D.~Talayco, A.~Vahdat, G.~Varghese, {\em et~al.}, ``P4:
  Programming protocol-independent packet processors,'' {\em ACM SIGCOMM
  Computer Communication Review}, vol.~44, no.~3, pp.~87--95, 2014.

\bibitem{shahbaz2016pisces}
M.~Shahbaz, S.~Choi, B.~Pfaff, C.~Kim, N.~Feamster, N.~McKeown, and J.~Rexford,
  ``Pisces: A programmable, protocol-independent software switch,'' in {\em
  Proceedings of the 2016 conference on ACM SIGCOMM 2016 Conference},
  pp.~525--538, ACM, 2016.

\bibitem{pfaff2015design}
B.~Pfaff, J.~Pettit, T.~Koponen, E.~J. Jackson, A.~Zhou, J.~Rajahalme,
  J.~Gross, A.~Wang, J.~Stringer, P.~Shelar, {\em et~al.}, ``The design and
  implementation of open vswitch.,'' in {\em NSDI}, pp.~117--130, 2015.

\bibitem{hwang2015netvm}
J.~Hwang, K.~Ramakrishnan, and T.~Wood, ``Netvm: high performance and flexible
  networking using virtualization on commodity platforms,'' {\em IEEE
  Transactions on Network and Service Management}, vol.~12, no.~1, pp.~34--47,
  2015.

\bibitem{ocp2015wedge}
Facebook and O.~C. Project, ``Facebook wedge - 16x40gb qsfp+ - leaf/spine
  switch,'' 2015.

\bibitem{intel2014data}
Intel, ``Data plane development kit,'' 2014.

\bibitem{att2016ecomp}
{\relax AT\&T inc.}, ``A{T}\&{T} ecomp (enhanced control, orchestration,
  management, and policy) architecture white paper,'' 2016.

\bibitem{han2015network}
B.~Han, V.~Gopalakrishnan, L.~Ji, and S.~Lee, ``Network function
  virtualization: Challenges and opportunities for innovations,'' {\em IEEE
  Communications Magazine}, vol.~53, no.~2, pp.~90--97, 2015.

\bibitem{att2013domain}
{\relax AT\&T}.~inc., ``A{T}\&{T} domain 2.0 vision white paper,'' 2013.

\bibitem{pci2010rev3}
{\relax PCI Special Interest Group}, ``{\relax PCI Express Base Specification
  Reveision 3.0},'' 2010.

\bibitem{mckeown1999achieving}
N.~McKeown, A.~Mekkittikul, V.~Anantharam, and J.~Walrand, ``Achieving 100\%
  throughput in an input-queued switch,'' {\em Communications, IEEE
  Transactions on}, vol.~47, no.~8, pp.~1260--1267, 1999.

\bibitem{west2001introduction}
D.~B. West {\em et~al.}, {\em Introduction to graph theory}, vol.~2.
\newblock Prentice hall Upper Saddle River, 2001.

\bibitem{sun2013fast}
W.~Sun and R.~Ricci, ``Fast and flexible: Parallel packet processing with gpus
  and click,'' in {\em Proceedings of the ninth ACM/IEEE symposium on
  Architectures for networking and communications systems}, pp.~25--36, IEEE
  Press, 2013.

\bibitem{smith2009evaluating}
R.~Smith, N.~Goyal, J.~Ormont, K.~Sankaralingam, and C.~Estan, ``Evaluating
  gpus for network packet signature matching,'' in {\em Performance Analysis of
  Systems and Software, 2009. ISPASS 2009. IEEE International Symposium on},
  pp.~175--184, IEEE, 2009.

\bibitem{nfp4000}
Netronome, ``Nfp-6000 intelligent ethernet controller family.''
  \url{https://www.netronome.com/static/app/img/products/silicon-solutions/PB_NFP6000.pdf}.

\bibitem{ozdag2012intel}
{\relax Intel FlexPipe}, ``Intel ethernet switch fm6000 series-software defined
  networking,'' 2012.

\bibitem{xpliant}
{\relax Cavium XPliant}, ``Xpliant packet architecture,'' 2014.

\bibitem{zilberman2014netfpga}
N.~Zilberman, Y.~Audzevich, G.~A. Covington, and A.~W. Moore, ``Netfpga sume:
  Toward 100 gbps as research commodity,'' {\em IEEE Micro}, vol.~34, no.~5,
  pp.~32--41, 2014.

\bibitem{pci2010sriov}
{\relax PCI Special Interest Group}, ``{\relax Single Root I/O Virtualization
  Revision 1.1},'' 2010.

\bibitem{jackson2012pci}
M.~Jackson and R.~Budruk, {\em PCI Express Technology: Comprehensive Guide to
  Generations 1.x, 2.x, 3.0}.
\newblock {\relax MindShare}, 2012.

\bibitem{sherwood2009flowvisor}
R.~Sherwood, G.~Gibb, K.-K. Yap, G.~Appenzeller, M.~Casado, N.~McKeown, and
  G.~Parulkar, ``Flowvisor: A network virtualization layer,'' {\em OpenFlow
  Switch Consortium, Tech. Rep}, pp.~1--13, 2009.

\bibitem{ek1999ieee}
N.~Ek, ``Ieee 802.1 p, q-qos on the mac level,'' {\em Apr}, vol.~24,
  pp.~0003--0006, 1999.

\bibitem{lameter2013numa}
C.~Lameter, ``Numa (non-uniform memory access): An overview,'' {\em Queue},
  vol.~11, no.~7, p.~40, 2013.

\bibitem{emmerich2015moongen}
P.~Emmerich, S.~Gallenm{\"u}ller, D.~Raumer, F.~Wohlfart, and G.~Carle,
  ``Moongen: a scriptable high-speed packet generator,'' in {\em Proceedings of
  the 2015 ACM Conference on Internet Measurement Conference}, pp.~275--287,
  ACM, 2015.

\bibitem{donard}
Microsemi, ``{\relax Project Donard: Peer-to-Peer Communication with NVM
  Express Devices},'' 2014.

\bibitem{directgpu}
Mellanox, ``{\relax Mellanox: NVIDIA GPU-Direct technology—accelerating
  GPU-based systems},'' 2010.

\bibitem{lu2011serverswitch}
G.~Lu, C.~Guo, Y.~Li, Z.~Zhou, T.~Yuan, H.~Wu, Y.~Xiong, R.~Gao, and Y.~Zhang,
  ``Serverswitch: A programmable and high performance platform for data center
  networks.,'' in {\em Nsdi}, vol.~11, pp.~2--2, 2011.

\bibitem{li2016clicknp}
B.~Li, K.~Tan, L.~L. Luo, Y.~Peng, R.~Luo, N.~Xu, Y.~Xiong, and P.~Cheng,
  ``Clicknp: Highly flexible and high-performance network processing with
  reconfigurable hardware,'' in {\em Proceedings of the 2016 conference on ACM
  SIGCOMM 2016 Conference}, pp.~1--14, ACM, 2016.

\bibitem{wood2015toward}
T.~Wood, K.~Ramakrishnan, J.~Hwang, G.~Liu, and W.~Zhang, ``Toward a
  software-based network: integrating software defined networking and network
  function virtualization,'' {\em IEEE Network}, vol.~29, no.~3, pp.~36--41,
  2015.

\bibitem{price2012opnfv}
C.~Price and S.~Rivera, ``Opnfv: An open platform to accelerate nfv,'' {\em
  White Paper. A Linux Foundation Collaborative Project}, 2012.

\bibitem{hancock2016hyper4}
D.~Hancock and J.~van~der Merwe, ``Hyper4: Using p4 to virtualize the
  programmable data plane,'' in {\em Proceedings of the 12th International on
  Conference on emerging Networking EXperiments and Technologies}, pp.~35--49,
  ACM, 2016.

\bibitem{cerrato2014supporting}
I.~Cerrato, M.~Annarumma, and F.~Risso, ``Supporting fine-grained network
  functions through intel dpdk,'' in {\em Software Defined Networks (EWSDN),
  2014 Third European Workshop on}, pp.~1--6, IEEE, 2014.

\bibitem{mckeown1999islip}
N.~McKeown, ``The islip scheduling algorithm for input-queued switches,'' {\em
  Networking, IEEE/ACM Transactions on}, vol.~7, no.~2, pp.~188--201, 1999.

\bibitem{karol1987input}
M.~Karol, M.~Hluchyj, and S.~Morgan, ``Input versus output queueing on a
  space-division packet switch,'' {\em IEEE Transactions on communications},
  vol.~35, no.~12, pp.~1347--1356, 1987.

\bibitem{dang2017whippersnapper}
H.~T. Dang, H.~Wang, T.~Jepsen, G.~Brebner, C.~Kim, J.~Rexford, R.~Soul{\'e},
  and H.~Weatherspoon, ``Whippersnapper: A p4 language benchmark suite,'' in
  {\em Proceedings of the Symposium on SDN Research}, pp.~95--101, ACM, 2017.

\end{thebibliography}
